\documentclass[fleqn,12pt,twoside]{article}
\usepackage{espcrc1}
\usepackage{epsfig}

\usepackage{graphicx}
\usepackage[figuresright]{rotating}

\newcommand{\AmS}{{\protect\the\textfont2
  A\kern-.1667em\lower.5ex\hbox{M}\kern-.125emS}}

\title{Stability of strange quark matter in the MIT bag model and in the Color Dielectric Model}

\author{C.Ratti\address[MCSD]{Dipartimento di Fisica Teorica, Universit\`a di Torino and INFN, Sezione di Torino, \\ 
        via P.Giuria 1, 10125 Torino, Italy}}%

\begin{document}

\maketitle

\begin{abstract}
We utilize the MIT bag model and two different versions of the Color Dielectric Model in order to study the properties of strange matter, and to discuss the stability of strangelets. We also investigate the effect of the introduction of perturbative gluons. The MIT bag model and the Double Minimum version of the Color Dielectric Model allow the existence of strangelets, while the Single Minimum version excludes this possibility.
\end{abstract}

\section{Strangelets in the MIT bag model}
We consider homogeneous quark matter made up of $u$, $d$ and $s$ quarks. 
We assume that, during a high energy collision between heavy ions, this state 
of matter, if formed, can only survive for a very short time, so that it 
has no time to reach $\beta$ equilibrium; hence we do not impose chemical 
equilibrium on the density of strange quarks, limiting 
 ourselves to assume that there exists in the system  a strange fraction 
$R_s=\rho_s/\rho$, $\rho$ being the total baryon density of quarks and 
$\rho_s$ the baryonic density of strange quarks.
Our aim is to find out whether this system is more or 
less stable than hyperons, in order to understand which state is more likely 
to be produced in heavy ion collisions, either hyperons and strange mesons or
 strangelets. For all the details see Ref.~\cite{Noi}.
We evaluate the minimal energy per baryon number of our system as a function of $R_s$ and compare this curve with the experimental masses of hyperons and the theoretical ones calculated in the model according to formula (3.6) of Ref.\cite{DeGrand75}.  We do not take into account surface effects, which would increase the energy of about 50-100 MeV.
We use different combinations of the model parameters, and in all cases we find that in the MIT bag model strange quark matter is more stable than hyperons. Even after the introduction of perturbative gluons, which contribution to the energy density is given by~\cite{Fahri84}:
\begin{eqnarray}
\epsilon_f^{OGE}&&=-\frac{\alpha_s}{\pi^3}m_f^4\left\{x_f^4-\frac{3}{2}
\left[\ln\left(\frac{x_f+\eta_f}{\eta_f}\right)-x_f\eta_f\right]^2+\right.
\nonumber\\
&&\qquad\left.+
\frac{3}{2}\ln^2\left(\frac{1}{\eta_f}\right)
-3\ln\left(\frac{\mu}{m_f\eta_f}\right)\left[\eta_fx_f
-\ln\left(x_f+\eta_f\right)\right]\right\}\,.
\label{ogemit}
\end{eqnarray}
where: 		$x_f=\frac{k_{F_f}}{m_f}$,	$\eta_f=\sqrt{1+x_f^2}$,
we find that the MIT bag model, apart from rather extreme choices of the model parameter, allows the existence of metastable strangelets.

\begin{figure}[t]
\psfig{file=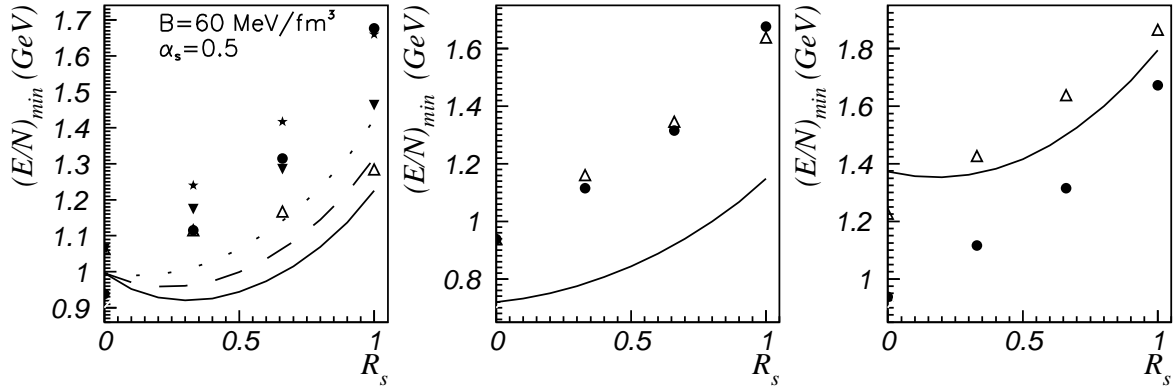,width=\textwidth}
\vspace {-1.5 cm}
\caption{Left panel: results in the MIT bag model. Middle panel: results in the DM version of the CDM. Right panel: results in the SM version of the CDM. In all panels: full circles are the experimental Hyperon masses, while all other symbols are the calculated masses: for details see Ref.~\cite{Noi}. }
\label{fig:largenenough}
\end{figure}

\section{Strangelets in the CDM.}
The Lagrangian of the model is:
\begin{eqnarray}
{\cal L} &&= \sum_{f=u,d,s}{\bar{\psi_f}} i\gamma^{\mu}\left(\partial_{\mu}
-ig_s\frac{\lambda^a}{2}A^a_{\mu}\right)\psi_f
-\frac{g f_{\pi}}{\chi}\sum_{f=u,d}{\bar{\psi_f}} \psi_f
-m_s\left(\chi\right) {\bar{\psi_s}} \psi_s +
\nonumber\\
&&\quad
+\frac{1}{2}\left(\partial_{\mu}\chi\right)^2-U\left(\chi\right)
-\frac{1}{4}\kappa\left(\chi\right)F^a_{\mu\nu}F^{a \mu\nu}\,,
\label{CDMlagr}
\end{eqnarray}
where $\psi_f$ are the quark fields, $A^a_{\mu}$ is the (effective) gluon 
field, $F^a_{\mu\nu}$ its strength tensor and $\chi$ is the color dielectric 
field, a scalar field which represents a multi-gluon state; we choose two different forms for the potential $U(\chi)$: a quadratic one (corresponding to a single minimum, SM) and a quartic one (corresponding to a double minimum, DM). We compare our results with the masses evaluated in the works by Aoki {\it et al.}~\cite{Aoki} (DM version) and by J. McGovern~\cite{Giuditta} (SM version). 

We find that in the DM version strangelets are more stable than hyperons, while in the SM version they are not, as shown in Fig.~\ref{fig:largenenough}. In the latter our curves are slightly below the calculated masses, and well above the experimental ones; thus, taking into account surface effects of about 100 MeV, we can conclude that in this case strangelets are forbidden. 

Our analysis shows that those models which describe a two-phase picture, like the MIT bag and the DM version of the CDM, allow the existence of strangelets, while the SM version of the CDM does not.

\end{document}